\documentclass[nologo,url,11pt,a4paper]{ETHpaper}
\usepackage{graphicx}
\usepackage{amsfonts}
\usepackage{amssymb}

\begin{document}

\title{Diversity-induced resonance in a model for opinion formation}
\author{Claudio J. Tessone and Ra\'ul Toral}
\address{$^1$ Chair of Systems Design, ETH Zurich, Kreuzplatz 5, CH-8032\\
$^2$ IFISC (Instituto de F{\'\i}sica Interdisciplinar y Sistemas Complejos),
CSIC-UIB, Ed. Mateu Orfila, Campus UIB, E-07122 Palma de Mallorca, Spain }
\reference{\emph{submitted}}

\maketitle
\abstract{ We study an opinion formation model that takes into account that individuals have diverse preferences when forming their opinion regarding a particular issue. We show that the system exhibits a phenomenon called ``diversity-induced resonance'' [Tessone et al. Phys. Rev. Lett. {\bf97}, 194101 (2006)], by which an external influence (for example advertising, or fashion trends) is better followed by populations having the right degree of diversity in their preferences, rather than others where the individuals are identical or have too different preferences. We support our findings by numerical simulations of the model and a mean-field type analytical theory.}

\date{\today}

\section{Introduction}
\label{introduction}

Resonance in forced dynamical systems is a topic of wide\-spread interest with many applications. A resonance is a maximum in a suitably defined response of the system to an external forcing. It usually requires the tuning of some system parameters to the optimal value. The simplest example is that of a forced linear oscillator whose amplitude of oscillations (the response) reaches a maximum when the natural frequency of the oscillator matches that of the external forcing. In coupled nonlinear systems, more complex resonances can appear. For example. they may mode-lock into states where the ratio between the individual frequencies are rational numbers in wide parameter regions, a phenomenon known as Arnold tongues \cite{SHS:1994}.

It was shown in the early 80's that a resonance can also appear as a function of the intensity of the fluctuations, of either internal or external origin. The basic mechanism leading to this {\sl stochastic resonance} \cite{BSV:1981,NN:1981} is rather generic and in its simplest form requires only a bistable system, a sub-threshold periodic forcing and a fluctuating additive term (noise) in the dynamics. The forcing induces a periodic lowering of the barrier separating the two stable fixed points, so helping the fluctuation dynamics to overcome the barrier in one direction or the other. The matching for resonance occurs when half the period of the external forcing equals the Kramers' time. The surprising result that fluctuations can enhance the response of a dynamical system to external forcing has become a new accepted paradigm and there have been many extensions and applications \cite{GHJM:1998}, including neural systems \cite{WPPDM:1994}, non-linear electronic devices \cite{FH:1983}, sensory systems \cite{DWPM:1993}, social dynamics \cite{KZ:2002,WW:1991}, etc.

Although most of the work in this field has considered simple, low-dimensional systems, more recent work analyses the role of fluctuations in the response of an {\sl extended} system \cite{LMDIB:1995,HSW:1996}. A usual modelling is that of many interacting units located on the sites of a lattice, such that the individual responses to the forcing are modified by the mutual interactions. A typical assumption in this case is that all the units are identical in the sense that they all possess the same values for all constituent parameters and that there is some regularity in the network of interactions.  For most applications, mostly in the biological or social sciences, those assumptions are not correct since some sort of diversity or variability will ineluctably be present. We have shown in a recent work \cite{TMTG:2006} the existence of a new type of resonance as a function of the diversity of the system. That work focuses on bistable and excitable systems in which the diversity is modelled by {\sl quenched noise} or, more specifically, by a parameter that adopts a different value for each of the units. Related work\cite{TSCT:2007} has shown that diversity can also induce synchronised spiking in an extended, unforced, excitable system.

Surprising at first, the fact that the right amount of diversity can enhance the response to an external forcing is not against our intuition. Think, for example, of a society which is very homogeneous in that all members of the population work on a particular economical field. If the economy tilts and that particular field becomes of less importance, it will have a big negative impact in the overall wealth of the population since individuals will not be able to follow the change. However, if there is some degree of heterogeneity and fractions of the populations work on different fields, there will be always a section that can adapt easily to the changing economy. The final ingredient that allows the whole society to follow the change is some degree of interaction by which the benefited agents can pull the others towards the new field.

These ideas were put forward in ref.~\cite{TMTG:2006} where we presented a mathematical model that displays this effect of diversity-induced resonance. We considered an ensemble of globally coupled $N$ bistable units $x_i(t)$ whose dynamics is given by:
\begin{equation}
\dot x_i=x_i-x_i^3+a_i+\frac{C}{N} \sum_{j=1}^N(x_j-x_i)+A \sin(\Omega t).
\end{equation}
The parameters $a_i$ are independently drawn from a Gaussian distribution of zero mean and variance $\sigma^2$. The value of $\sigma$ can be considered as a measure of the diversity. If $\sigma=0$, all systems are identical, whereas increasing values for $\sigma$ indicate a larger degree of heterogeneity. We consider that the external periodic forcing of amplitude $A$ is sub-threshold for those systems with $a_i$ close to zero. This means that for $\sigma=0$ the system as a whole is unable to display a wide response to the forcing and the collective variable $X(t)=\frac{1}{N}\sum_{i=1}^Nx_i(t)$ oscillates around one of the equilibrium values $X(t)=\pm 1$ with an oscillation amplitude proportional to $A$.  Imagine that the oscillation point is $X(t)=+1$. As $\sigma$ increases there will be a fraction of units (those with a sufficiently large, negative, value for $a_i$) for which the weak forcing is now sufficient to take them to the minimum $x_i=-1$. If the intensity of the coupling $C$ is sufficiently large the whole system will be pulled by those units and taken to that minimum. Hence, the collective variable will have performed a large excursion from $X(t)=+1$ to $X(t)=-1$. The opposite jump from $X(t)=-1$ to $X(t)=+1$ is induced when the sign of the forcing is reversed as induced by those units that have a large, positive, value for $a_i$.

It is important to realise, as explained in the theoretical treatment of \cite{TTV:2007}, that this resonance mechanism relies of the individual units having different dynamical response to the external forcing but that the origin of the heterogeneity in the response is not important. Different sources of disorder such as noise, diversity, non-regular network of connectivities, inhibitory couplings, etc. can be the origin of the resonance. The combined effect of noise and diversity has been analysed in \cite{gassel2007,GGK:2008}. This result has already been shown of interest in many fields, from complex networks \cite{acebron2007} to cellular signalling \cite{chen2007}. A linear model in which a full analytical calculation is possible has been recently studied\cite{THG:2008}.

In this paper, we show a rather different example of diversity-induced resonance in an opinion formation model. The model is a simple majority model with the addition of preferences in the individual choices. Those preferences vary from individual to individual and are the source of diversity. The interest of the paper is twofold. First, by giving an example which is very far away from the dynamical system described above, we want to emphasise the generality of the mechanism leading to the resonance. Second, we believe that the example has interest on its own in the field of social sciences, since it shows that an external forcing (imitating the effect of advertising) has a larger impact on a heterogeneous society than on a completely homogeneous one. This effect might be relevant when explaining the changes in opinion (e.g. in poll's results) motivated by an apparently small change in the external environment.

The outline of the paper is as follows: in section \ref{model} we define the model for opinion formation and highlight its formal similarities to other well known models of spin-glass systems, while stressing the ingredients that, according to the general discussion, might lead to a resonance effect. In section \ref{results} we present the results of numerical simulations that show the existence of the resonance as a function of a parameter measuring the diversity in the individual preferences. In section \ref{theory} we introduce a  mean-field theory that focuses on the collective variable and from which a global mechanism for the resonance can be extracted. Finally, in section \ref{conclusions} we end with brief conclusions and outlooks.

\section{Model studied} 
\label{model}

Although the focus is very different, the model we have introduced bears many similarities with the random field Ising model~\cite{SDP:2004}. This model has attracted much attention because of its interest for modelling disordered magnetic materials \cite{shinbrot2001} and also spin-glasses \cite{weissman1993}. More recent work has focused on the hysteresis behaviour when subjected to a slowly varying magnetic field \cite{IRSV:2006} or the question on how the system can reach the global energy minimum \cite{sarjala2006}. The use of techniques of statistical physics to model social behaviour has a long tradition (see, e.g., \cite{WW:1991})  and suitable modifications of the random field Ising model have been used already to model the dynamics of social systems. To the best of our knowledge, Galam~\cite{galam:1997} was the first one to model individual preferences by a random field. Stochastic resonance induced by fluctuating terms in the dynamics was described by Kuperman and Zanette \cite{KZ:2002}. Michard and Bouchaud \cite{michard2005} used the random field Ising model in an external field and studied the emergence of collective opinion shifts in a diverse population. The imitation mechanism, included in this work,  has also been presented as a key feature in price formation dynamics by Zhou and Sornette~\cite{zhou2007}.

We consider a  model for opinion formation in which the opinion on a particular topic is considered to be a binary variable (against or in favour of such a topic). There are $N$ individuals, each one having an opinion $\mu_i(t)=\pm 1$, $i=1,\dots,N$, at time $t$. The opinion $\mu_i(t)$ of individual $i$  can change due to (i) the interaction with the $k_i$ individuals in its neighbourhood $n(i)$, modelled by a  majority rule, and (ii) the influence of advertising, modelled as the effect of some external time-varying agent. 
Each individual has a tendency to favour one of the two opinions, $+1$ or $-1$. We introduce diversity in the fact that this preference for one of the two opinions is stronger in some individuals than in others. We model the effect of individual preferences by a set of independent parameters $\theta_i$. They are drawn from a probability distribution $g(\theta)$, which satisfies $\langle \theta_i \rangle = 0$, $\langle \theta_i\, \theta_j \rangle = \delta_{ij}\,\sigma^2$. Their influence becomes apparent when we spell out the evolution rules of this model:
\begin{enumerate}
 \item[(i)] Select randomly one individual $i$. Its opinion at time $t$ is modified as :
\begin{equation}
\label{evol1}
\mu_i(t+dt)=\mathrm{sign} \left[ \frac 1 k_i \sum_{j\in n(i)}\mu_j(t) + \theta_i
\right].
\end{equation}
In words, individual $i$ adopts the average opinion in its neighbourhood when this average opinion overcomes its preference $\theta_i$. This is a mechanism of social pressure weighted against individual preferences. For instance if $\theta_i=0.3$ (resp. $\theta_i=-0.3$) it is necessary that the proportion of neighbours supporting the $-1$ (resp. $+1$) opinion exceeds $70\%$ in order for individual $i$ to adopt the majority opinion. Note that when $|\theta_i|>1$ the individual will keep its preferred opinion no matter what the social pressure is.
\item[(ii)]   With probability $A| \sin(\Omega t )  |$, the opinion is set to 
\begin{equation}
\label{evol2}
\mu_i(t+dt)=\mathrm{sign}\left[\sin(\Omega t)+ \alpha\, \theta_i \right].
\end{equation}
This represents the effect of a time dependent external global field (advertising, for example). $A$ is a measure of the strength of the field. This field has to overcome the preference $\theta_i$ (weighted by a scale factor $\alpha$) in order for the individual to adopt the value favoured by the field,  i.e.: the advertising has more effect on those individuals whose preference coincides with the sign of the advertising.
\end{enumerate}
After these two steps have been taken, time increases by $t\to t+d t=t+1/N$ and a new individual is selected again at random. The process is repeated for many cycles of the external forcing. We are interested in quantifying how well the system globally responds to the external forcing. To this end, we focus on the time evolution of the average opinion:
\begin{equation}
m(t) =\frac 1 N \sum_{i=1}^N \mu_i(t).
\end{equation}
In general, $m(t)$ oscillates in time with the frequency $\Omega$ of the forcing. The amplitude of the oscillations of $m(t)$ is a measure of the response to the forcing. An equivalent measure, but  more useful from the computational point of view, is the so-called spectral amplification factor $R$ defined as\cite{GHJM:1998}: 
\begin{equation}\label{eq:saf}
R=4 A^{-2}|\langle m(t) {\rm e}^{-i\Omega t}\rangle |^2, 
\end{equation}
where $\langle \dots \rangle$ denotes a time average. 

The main result of this paper, as shown in the next sections, is the existence of a value of the diversity $\sigma$ for which the response $R$ takes a maximum value. This resonance effect appears for weak forcing ($A$ sufficiently small) and implies that the advertising has an optimal effect on the population when there is some degree of diversity in the preferences. The ``microscopic'' mechanism is easy to understand. In a diverse society, there is always a fraction of the population which is receptive to follow the external field. This fraction initiates a change in the opinion and then, by the imitation mechanism, the change is spread towards a larger fraction. If the population is not too diverse, the fraction that can follow the external signal is small and it is not enough to initiate the global change. If the population is too diverse, however, the imitation mechanism is not effective. In section \ref{theory} we will present a  mean-field theory that offers an alternative, ``macroscopic'', explanation to this microscopic mechanism. It will be clear after the theoretical treatment that the origin of the resonance can be traced to the lack of order caused by the diversity and, as stressed in \cite{TMTG:2006}, any source of disorder will lead to similar results. A recent analysis of a similar model\cite{MTS:2008} shows that the  
disorder caused by competitive interactions also leads to a resonance effect.

\section{Numerical results}
\label{results}

\begin{figure}[t!]
\begin{center}
\includegraphics[width=7cm]{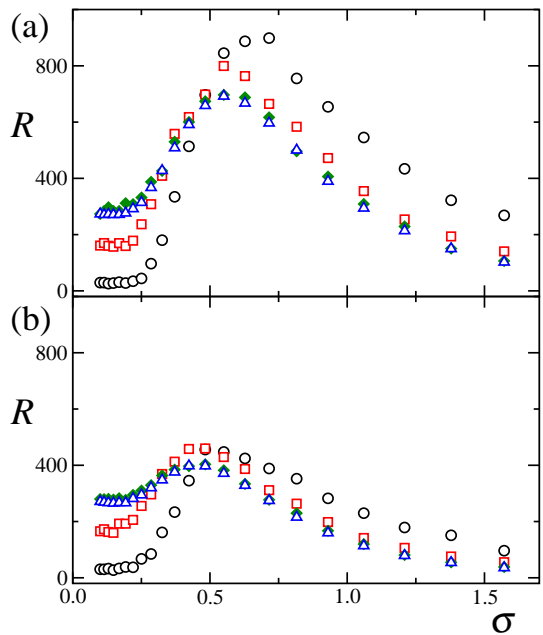}
\end{center}
\caption{\label{opn:saf-sigma-2d} Spectral amplification factor, $R$, as a function of the diversity $\sigma$ when the neighbourhood network is a two-dimensional regular lattice. Each panel correspond to different values of the parameter $\alpha$, that measures the relative weight of the individual preferences with respect to the external signal: (a) $\alpha = 0$, (b) $\alpha = 1$. The symbols represent different system sizes: $N=5^2$ ($\circ$), $N=10^2$ ($\Box$), $N=30^2$ ($\diamond$) and $N=100^2$ ($\vartriangle$). All the curves show an optimum response for an intermediate value of diversity. In both panels, the signal has an intensity $A=5\times 10^{-3}$ and frequency $\Omega=2\pi/1024$.
}
\end{figure}

We have ran the model using different topologies for the neighbourhood network, namely: a two-dimensional lattice with von Neumann neighbourhood (and periodic boundary conditions); a fully-connected network; and a small-world network \cite{WS:1998}. In each case, we set the amplitude of the forcing $A$ to a small, sub-threshold, value such that, in the absence of diversity, the average opinion makes small oscillations around the value $m=+1$ or $m=-1$. As the diversity increases, the amplitude of the oscillations first increases and then decreases again. 

For the two-dimensional lattice, we plot in figure \ref{opn:saf-sigma-2d} the spectral amplification factor $R$ as a function of diversity $\sigma$ for two different values of the parameter $\alpha$, which determines the relative importance that preference has with respect to the external signal as compared to the neighbours influence. Although there are some clear finite-size effects, a well defined maximum is clearly observable in both cases. Comparing panels \ref{opn:saf-sigma-2d}(a) and \ref{opn:saf-sigma-2d}(b), it is apparent that for $\alpha=1$ the response at the optimal diversity level is lower that for $\alpha=0$. The reason for this dependency on the parameter $\alpha$ can be easily understood by inspection of the dynamic rules of the system (cf.~Eq.~\ref{evol2}): this parameter can be seen as a secondary source of disorder in the system disturbing the external signal, lowering its effect. Thus, the effective level of disorder introduced by the diversity in the population is increased, for increasing values of $\alpha$. This can be confirmed by the fact that in panel (b), the location of the maximum response is shifted towards lower values of diversity, measured in terms of $\sigma$.

\begin{figure}[t!]
\begin{center}
\includegraphics[width=9cm]{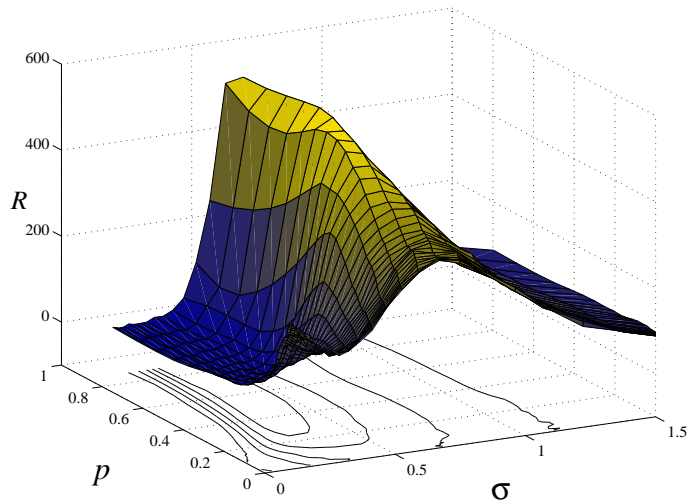}
\end{center}
\caption{\label{opn:saf-sigma-surfsw} For a small-world neighbourhood network,
we plot the response $R$ as a function of diversity, $\sigma$, and the
rewiring probability $p$. It is
apparent that, regardless the exact network topology, the effect of
diversity-induced resonance appears in the system, showing the existence of an
optimum synchronisation between the external signal and the global dynamics of
the system. The signal
has an intensity $A=5\times 10^{-3}$ and frequency $\Omega=2\pi/1024$. The
system size is $N=10^3$. The initial network is a one-dimensional one with
$k=3$, i.e.~before rewiring, each site is connected to its six nearest
neighbours. }
\end{figure}

Figure \ref{opn:saf-sigma-surfsw} shows the results for the small-world network. This is constructed in the usual way \cite{WS:1998}, with a rewiring probability $p$ and average connectivity $2k$. The main result here is that as $p$ increases (leading to a larger degree of disorder), the resonance peak narrows and the optimal response increases. Thus, for slightly disordered networks the systems reacts more robustly amplifying the external stimulus for a wider range of diversity values; however, $R$ reaches lower values. This effect can be understood if we consider that for small values of $p$, it is more probable for the system to develop stable domains of different opinions, that enlarge or shrink depending on the instantaneous signal value. These domains tend to coexist, thus decreasing the global response of the system.

\begin{figure}[t!]
\begin{center}
\includegraphics[width=7cm]{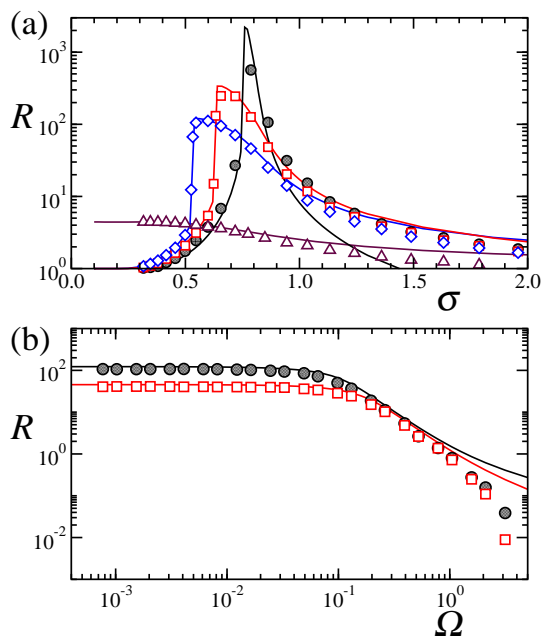}
\end{center}
\caption{\label{saf-mf} We compare the theoretical prediction of the
system response (see inline text for details) with numerical simulations
in fully-connected networks. In panel (a) each symbol represents the
response $R$, as a function of diversity $\sigma$ for different values
of signal intensity: $A=0.01$ ($\bullet$), $A=0.05$ ($\Box$),
$A=0.10$ ($\diamond$) and $A=0.60$ ($\vartriangle$). The full lines correspond
to the analytical prediction. The other parameters are: $N=10^4$,
$\Omega=2\pi/1024$ and $\alpha=0.1$.  Panel (b) shows the spectral
amplification factor as a function of signal frequency, $\Omega$ for two
different values of the signal intensity: $A=0.05$ ($\bullet$) and $A=0.1$
($\Box$). The solid lines stand for the theoretical results. The other
parameters are $\sigma=0.85$, $\alpha = 0.1$ and $N=10^4$. }
\end{figure}

Finally, figure \ref{saf-mf} shows the response of the system for a fully-connected neighbourhood network. In panel (a), different symbols correspond to different signal amplitudes. We have also included in this figure the case of a signal amplitude $A=0.60$ which is supra-threshold in the zero-diversity case.  As seen in this panel, supra-threshold signals are not amplified at all, and the response of the system monotonically decreases with diversity. This trait was found in diversity-induced resonance in bistable systems \cite{TMTG:2006}, but is also commonly found in systems exhibiting stochastic resonance \cite{GHJM:1998} with respect to noise intensity. Also, in the same panel, it can be seen that  larger (but still sub-threshold) signal amplitudes, lead to lower optimal diversity values. In panel (b) of figure \ref{saf-mf} we plot the dependence of the system response $R$ with respect to the signal frequency $\Omega$. As shown, the larger the frequency, the lower the response of the system. In this discrete model, this is because slower signals allow units with a given bias more akin to adopt its favoured opinion.

\section{Analytical approach}
\label{theory}

\begin{figure}[t!]
\begin{center}
\includegraphics[width=7cm]{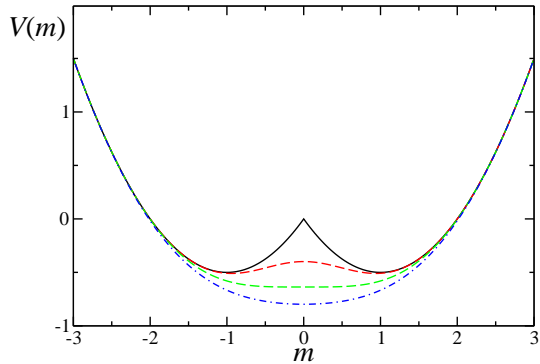}
\end{center}
\caption{\label{potential} Effective potential (cf.~Eq.~\ref{eq:potential}) defining the relaxational dynamics of the average opinion for different values of diversity: $\sigma=0$, $0.5$, $0.7979$, and $1$ (respectively, from top to bottom as found at the vertical line $m=0$).}
\end{figure}

We now present a  mean-field like theory that can explain the observed features. The derivation follows the lines of \cite{MTS:2008}. Since the opinion changes at each time by the modification of a single variable, we can write the following exact relation for the ensemble average $m(t)$:
\begin{equation}
Nm(t+dt)=N m(t)+\left\langle \mu_i(t+dt)-\mu_i(t)|\{\mu(t)\}\right\rangle
\end{equation}
where $\{\mu(t)\}=(\mu_1(t),\dots,\mu_N(t))$ denotes the particular realisation of the $\mu_i$ variables and $\langle\dots|\dots\rangle$ denotes a conditional ensemble average. By identifying $dt=1/N$ and rearranging we get:
\begin{equation}
\frac{dm(t)}{dt}=-m+\left\langle \mu_i(t+dt)|\{\mu(t)\}\right\rangle
\end{equation}
The average term in the right-hand side can be computed using the contribution from the biased majority rule, Eq.~(\ref{evol1}) which acts with probability $1-|f(t)|$  (we define $f(t)=A\sin(\Omega t)$) and  the contribution of the external forcing, Eq.~(\ref{evol2}) which acts with probability $|f(t)|$. In the spirit of the mean-field approximation we replace in Eq.~(\ref{evol1}) the average opinion of the neighbourhood $n(i)$ by the global average opinion $m(t)$. This yields:
\begin{equation}
\label{eq:mt}
\begin{array}{rcl}
\left\langle \mu_i(t+dt)|\{\mu(t)\}\right\rangle & =& \left(1-|f(t)|\right)\langle\mathrm{sign} \left( m(t) + \theta_i \right)|\{\mu(t)\}\rangle+ \\
& & |f(t)| \langle\mathrm{sign} \left(f(t) + \alpha \theta_i \right) |\{\mu(t)\}\rangle
\end{array}
\end{equation}
Both mean values can now be easily evaluated:
\begin{eqnarray}
\label{eq:meta}
\langle \mathrm{sign}(m(t) + \theta_i) \rangle  &=&  \mathrm{Prob}(\theta_i > -m(t) ) - \mathrm{Prob}(\theta_i <-m(t) ) \nonumber\\
&=& \hat G( m(t) ).
\end{eqnarray}
Here, $\hat G(\theta)=1-2G(-\theta)$, where $G(\theta)$ is the cumulative probability function of the distribution of preferences $g(\theta)$. In the same way, the contribution of the external signal to equation (\ref{eq:mt}) is
\begin{equation}\label{eq:feta}
\langle \mathrm{sign}(f(t) + \alpha\, \theta_i) \rangle  = \hat G(f(t) / \alpha ).
\end{equation}
Adding up those contributions we get a closed evolution equation for the average opinion $m(t)$:
\begin{equation}
\label{eq:mf:m}
\frac{dm(t)}{dt} =  - m + |f(t)|\,\hat G \left(f(t)/\alpha\right) + (1-|f(t)|)\, \hat G(m). 
\end{equation}
This equation can be written as a relaxational dynamics \cite{sMT:1999} in a time-dependent potential
\begin{equation}
\frac{dm(t)}{dt} =  - \frac{\partial V(m,t)}{\partial m}.
\end{equation}
It is easy now to see the effect that the diversity has on the dynamics of the global variable. For the sake of concreteness, we consider that the preferences follow a  Gaussian distribution of zero mean and variance $\sigma^2$, but similar results hold for other distributions. In this case, $\hat G(\theta)=\mathrm{erf}(\theta/\sigma\sqrt{2})$, being $\mathrm{erf}(x)$ the error function \cite{AS:1964}. Consider first the non-forced case, $f(t)=0$. The potential is:
\begin{equation}\label{eq:potential}
V(m)=\frac{m^2}{2} -m\,\mathrm{erf}\left(\frac{m}{\sigma\sqrt{2}}\right)-\sigma\sqrt{\frac{2}{\pi}}e^{-m^2/2\sigma^2}.
\end{equation}
This potential is bistable when $\sigma=0$. As $\sigma$ increases, the two minima of the potential get closer to each other and the barrier between them decreases until at the critical value $\sigma_c=\sqrt{2/\pi}$ the potential becomes monostable, as can be seen in figure \ref{potential}. This shows the existence of a phase transition between consensus and non-consensus states. The effect of the external field now is easily understood as a periodic lowering and rising of the two potential wells. For small $\sigma$, the barrier separating the two stable points is large and the effect of the field is that of making the global variable $m(t)$ oscillate around one of the equilibrium points. As $\sigma$ increases, the barrier lowers and it is possible to induce transitions between the two stable states. When $\sigma$ increases even further, the potential becomes monostable and again the effect of the forcing is that of producing small oscillations, this time around the only equilibrium point. This macroscopic mechanism is similar to the one found in \cite{TMTG:2006},

We can now integrate numerically equation \ref{eq:mf:m} to obtain the time evolution of $m(t)$ and compute the spectral amplification factor (cf.~Eq.~\ref{eq:saf}) for a given set of parameters. In both panels of figure \ref{saf-mf}, we display with lines this theoretical prediction. As expected from a mean-field type theory, there is a good agreement with the numerical results of the fully connected network.

\section{Conclusions}
\label{conclusions}

We have analysed theoretically and numerically a model for opinion formation. The model has many points in common with random field Ising models used to study phase transitions in statistical mechanics. It incorporates two basic ingredients for the evolution of the opinion held by an individual: social pressure and the effect of advertising (modelled as an oscillating influence acting over all the individuals). The model also considers that every individual has an intrinsic preference for one or the other option. We have shown that an optimal synchronisation of the average opinion with respect to the external signal can be achieved if the population shows some degree of diversity in the preferred opinions. We have also shown that the results are robust against the exact topology of the network used to model the neighbourhood of the individuals, and that the results hold for increasingly large system sizes. We have given explanations for this resonance both from the point of view of the individual responses to the external influence or by looking and the average global variable within a mean-field approach. 

From the point of view of the social dynamics, our results imply that an external message can propagate better in a society if there is some degree of diversity in the individual preferences. These results can also be interpreted in the context of population dynamics where the external signal stands for a changing environment \cite{droz2005,bena2007}. The value of the diversity parameter specifies to which external condition an individual is best fitted to. In this setting, the response is directly related to the average fitting of the population. Within this interpretation, the results reported in this paper imply that intermediate values of diversity cause a better fit of the population to the changing environment.

{\bf Acknowledgements}: We acknowledge financial support by the MEC (Spain) and FEDER (EU) through project FIS2007-60327. CJT acknowledges financial support from SBF (Swiss Confederation) through research project C05.0148 (Physics of Risk).

\end{document}